\documentclass[english,twocolumn,preprintnumbers,amsmath,amssymb,showpacs]{revtex4}
\pdfoutput=1
\usepackage{graphicx}
\usepackage{textcomp}
\usepackage{amsmath}
\usepackage{epsfig}
\usepackage{dcolumn}
\usepackage{rotating}
\usepackage{bm}
\input{epsf}
\usepackage{color}
\usepackage{graphicx}
\usepackage{mathrsfs}
\usepackage{theorem}
\usepackage{amsmath,amssymb}
\usepackage{times}

\usepackage{multirow}

\begin{document}

% Include your paper's title here

\title{A superconducting cavity bus for single Nitrogen Vacancy defect centres in diamond}

% Place the author information here.  Please hand-code the contact
% information and notecalls; do *not* use \footnote commands.  Let the
% author contact information appear immediately below the author names
% as shown.  We would also prefer that you don't change the type-size
% settings shown here.

\author{J. Twamley and S. D. Barrett}
\affiliation{Centre for Quantum Computer Technology, Macquarie University, Sydney, NSW 2109, Australia}
\date{\today}

\begin{abstract}
Circuit-QED has demonstrated very strong coupling between individual microwave photons trapped in a superconducting coplanar resonator and nearby superconducting qubits  \cite{Blais:2004p3701,Wallraff:2004p6255}. With the recent demonstration of a two-qubit quantum algorithm \cite{DiCarlo:2009p9586}, Circuit-QED has the potential to engineer larger quantum devices. However difficulties associated with performing single shot readout of the Circuit-QED qubits  and their short decoherence times are obstacles towards building large scale devices. To overcome the latter, hybrid designs have been proposed where one couples ensembles of polar molecules \cite{Andre:2006p9477, Rabl:2006p7458,Tordrup:2008p10034}, neutral atoms \cite{Verdu:2009p10030,Petrosyan:2008p10039}, Rydberg atoms \cite{Petrosyan:2009p9315},  Nitrogen-Vacancy centres in diamond \cite{Imamoglu:2009p9319}, or electron spins \cite{Wesenberg:2009p10065}, to the superconducting resonator. Rather than build a long lived quantum memory via coupling to an ensemble of systems we show how, by designing a novel interconnect, one can strongly connect the superconducting resonator, via a magnetic interaction, to a single electronic spin. By choosing the electronic spin to be within a Nitrogen Vacancy centre in diamond one can perform optical readout, polarization and control of this electron spin using microwave and radio frequency irradiation \cite{GurudevDutt:2007p9842,Neumann:2008p6488, Balasubramanian:2009p9876}. More importantly, by utilising Nitrogen Vacancy centres with nearby ${}^{13}C$ nuclei, using this interconnect, one has the potential build a quantum device where the nuclear spin qubits are connected over centimeter distances via the Nitrogen Vacancy electronic spins interacting through the superconducting bus.
\end{abstract}
\pacs{42.50.Pq, 03.65.-w, 03.67.-a, 37.30.+i}
\maketitle

Achieving strong coupling between light and matter plays a vital role in the study of strongly correlated quantum dynamics. When strong coupling is achieved the matter portions of the hybrid light-matter system may act as a quantum memory and this can play an essential role in a quantum repeater or a quantum computer architecture. If the matter systems have long decoherence times and can be easily initialised, controlled and selectively coupled/decoupled to the light bus, then one has the potential for the design of a scalable quantum device. Strong coupling is achieved when the coupling strength between the light and matter exceeds their respective decay rates $g>\kappa,\gamma$, and this means that the associated vacuum splittings can be resolved in a spectroscopic experiment. A number of theoretical proposals have been recently advanced for a hybrid Circuit-QED/atomic system where one couples the microwave photons trapped in the superconducting cavity to a nearby ensemble of atoms or molecules, held in a microtrap above the surface of the superconducting chip. Though technically demanding this type of design has the advantage of magnifying the light-matter coupling strength for an ensemble consisting of $N$ atomic/molecular systems by a factor $\sqrt{N}\times$ the individual system light-matter coupling  strength. A more convenient approach would be to couple to an ensemble of ``atomic-like'' solid state systems which have long decoherence times \cite{Wesenberg:2009p10065}. However as all such solid-state systems are not identical, the coupling to such an inhomogenous ensemble would suffer and, further, if the coupling is orientation dependent (magnetic dipole-dipole), then any misalignments of the individual ensemble dipoles could greatly diminish the light-matter coupling strength. One promising solid-state ``atomic like'' system which can couple magnetically to a superconducting cavity is a Nitrogen Vacancy defect in diamond \cite{Imamoglu:2009p9319}. However if the ensemble consists of NV containing diamond nanocrystals, then the NV dipoles are aligned randomly in space and the coupling to the superconducting cavity averages out. If the NVs are implanted into bulk diamond then the NV magnetic dipoles are randomly aligned along four possible orientations in the diamond lattice and their combined coupling to the superconducting cavity may be non-vanishing but could be significantly smaller than if they were all aligned. These complications lead one to consider the possibility of coupling just a single NV defect to the superconducting cavity. In what follows we will quickly determine that the light-matter coupling strength between light trapped in the best superconducting coplanar waveguide cavities fabricated to date, and the electron spin in a nearby single NV defect is an order of magnitude smaller than the cavity linewidth and thus this hybrid system is not strongly coupled. However, we find that by encircling the NV defect with a Persistent Current Qubit (PCQ) loop - or interconnect, we can achieve strong coupling between the electron spins in the NV and the photons in the coplanar resonator, i.e. the PCQ loop couples strongly to the coplanar resonator, while the persistent currents in the loop generate a large enough magnetic field at the centre of the loop to shift the resonance frequency of the NV microwave transitions by amounts larger than the cavity linewidth. We finally show that by adapting the traditional single-loop persistent current qubit to a multi-loop design one can magnify the NV-resonator effective coupling strength by the number of turns in the multiloop design. 
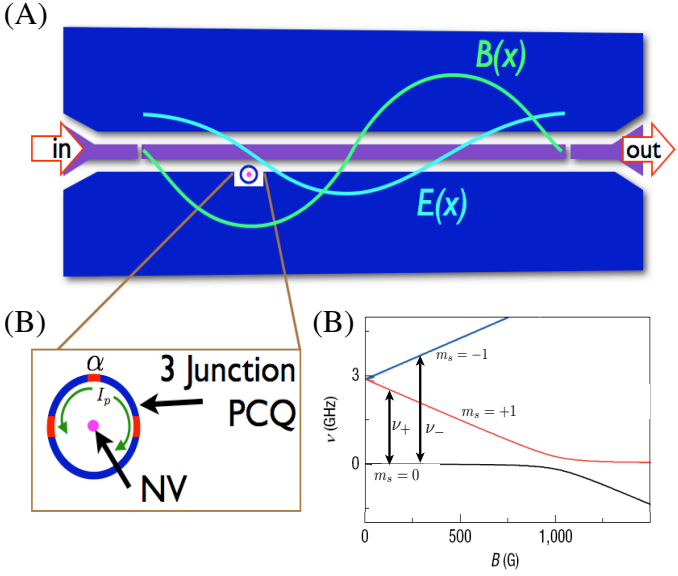
\begin{figure}[h]
\begin{center}
\setlength{\unitlength}{1cm}
\begin{picture}(7,7)
\put(-0.8,0){\includegraphics[width=8.8cm,height=6.6cm]{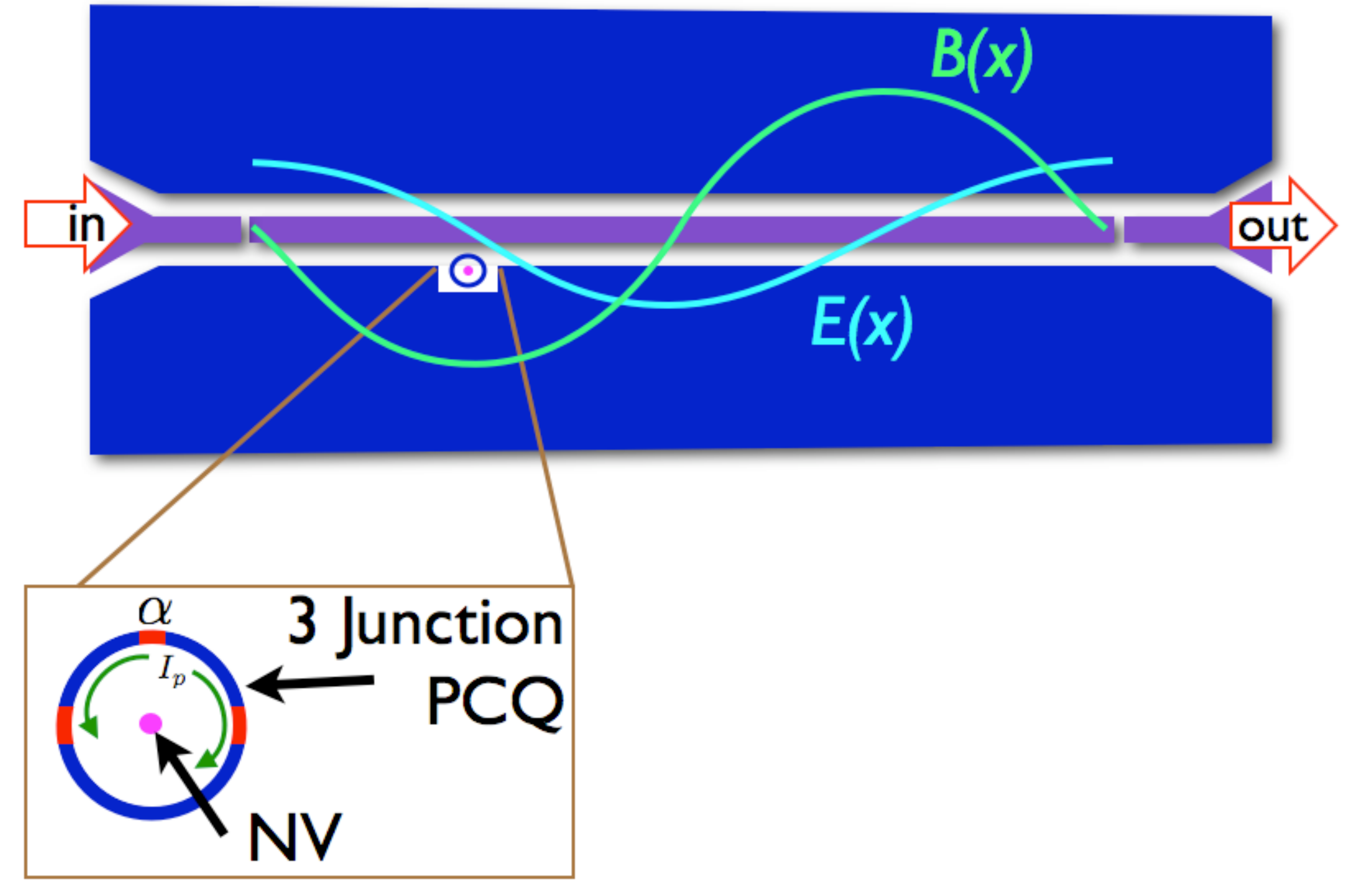}}
\put(3.2,-1){\includegraphics[width=4.5cm,height=3.9cm]{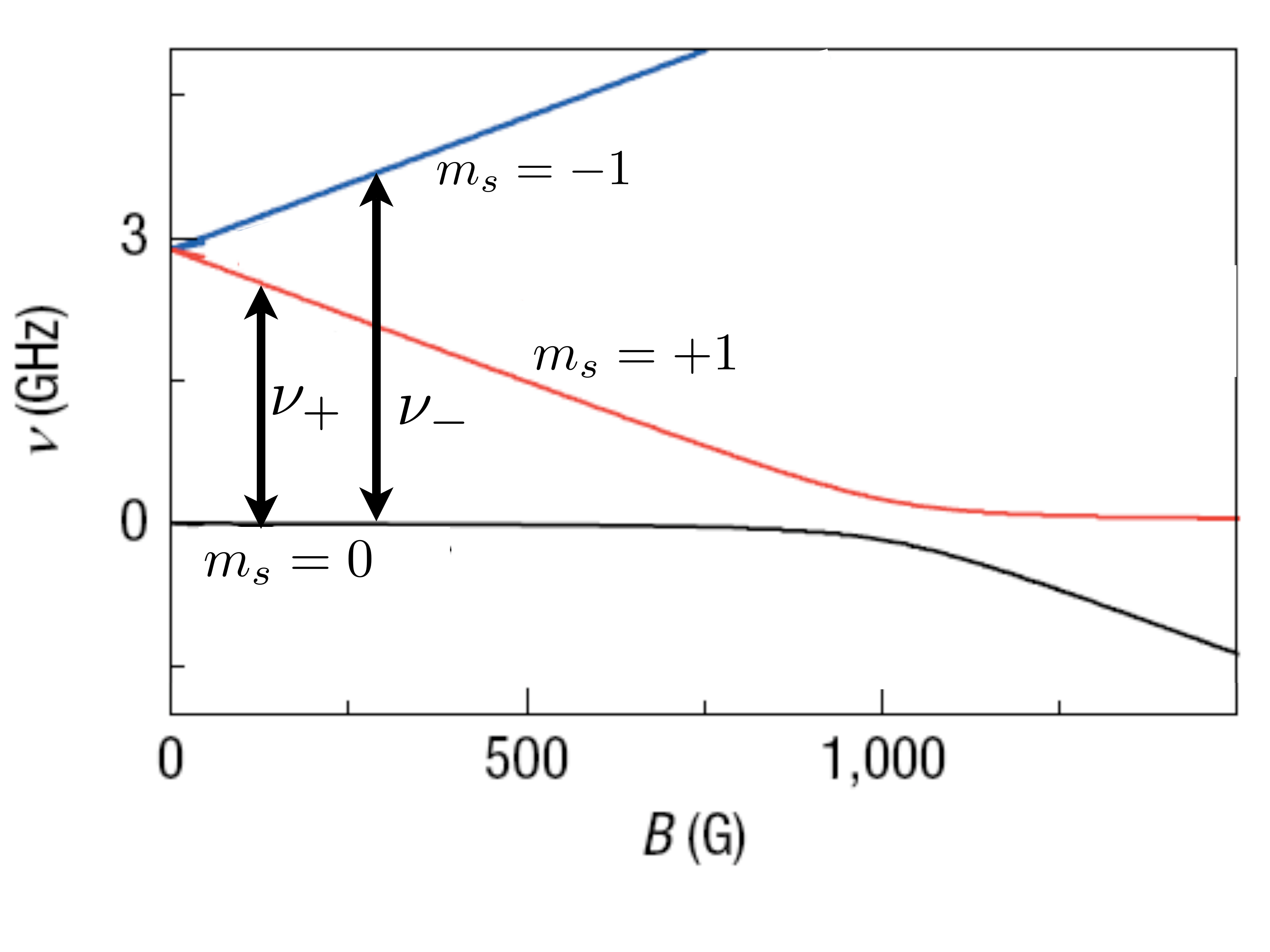}}
\put(-1,6.6){\large (A)}
\put(-1,2.5){\large (B)}
\put(3.1,2.5){\large (B)}
\end{picture}
\end{center}
	\caption{\label{PCQ}\textbf{Persistent current qubit (PCQ) loop interconnect}. (A) superconducting coplanar resonator with PCQ loop located at an antinode of $B(x)$ of the resonator. The PCQ loop encircles an individual magnetic spin system - in this case a Nitrogen-Vacancy defect in diamond. The PCQ loop couples via mutual inductance to the coplanar resonator and couples to the magnetic spin via the B-field at the centre of the loop generated by the persistent currents in the loop. 
		(B) Detail of the PCQ made up of three Josephson junctions (red), two identical and the other smaller by a factor $\alpha$, encircling the magnetic spin system (magenta), coupled via the persistent circulating currents $I_p$ (green). (C) Energy levels of ground state triplet ${}^3A_2$, of the NV as a function of applied magnetic field. }
\end{figure}

%\subsection{Coplanar waveguide resonator}
{\em Coplanar waveguide resonator:-} We consider a coplanar waveguide resonator (CPW), and the magnetic coupling between such a resonator and a nearby magnetic spin system. 
The Hamiltonian for microwave photons in a CPW resonator is $\hat{H}_{r}=\hbar \omega_r(\hat{a}^\dagger \hat{a}+1/2)$, and recent devices \cite{Niemczyk:2009p7767}, have reported $\omega_r/2\pi\sim 6$ GHz with a quality factor $Q\sim 2.3\times 10^5$, giving a cavity decay rate of $\kappa/2\pi\sim 26$ kHz. The total equivalent inductance of these resonators near their resonant frequency is typically a few nanoHenry $L_r\sim 2$ nH. We now estimate the size of the magnetic field generated by the vacuum fluctuations of the photons within the resonator. This will be used to estimate the size of the coupling directly to the NV when placed next to the central conductor of the resonator. The RMS current flowing through the resonator when the photon mode is in the ground state can be estimated to be $I_{rms}=\sqrt{\hbar\omega_r/2L_r}$. Assuming that in the superconducting state that the current in the central conductor flows in a thin strip at the surface we can estimate the magnetic field strength a distance $d$ away (see Fig {\ref{Comparison}), to be 
\begin{equation}
B_{0,rms}(d)=\mu_0I_{rms}/\pi d\;\;.
\label{B0-CPW}
\end{equation}

%\subsection{Coupling of the NV directly to the Coplanar Resonator}
{\em Coupling of the NV directly to the Coplanar Resonator:-} To estimate the size of the magnetic coupling between the electrons in the NV and a nearby CPW we take, for simplicity, the NV dipole axis to be along the $\hat{z}$ direction and describe the Hamiltonian for the ground state triplet (spin 1), ${}^3A_2$ electronic system of the NV by
\begin{equation}
\hat{H}_{NV}/\hbar=g_e\beta_eB_z\hat{S}_z+D\left(\hat{S}_z^2-\frac{2}{3}\mathbb{I}\right)\;\;,
\label{NV-Ham}
\end{equation}
where in the first Zeeman term $B_z$ is the $z-$component of the magnetic field at the NV, $\hat{S}_z$ is the $z-$spin 1 operator, $g_e=-2$, and $\beta_e/2\pi\sim 1.4\times 10^4$MHz/T. The second term is the so-called zero field splitting term with $D/2\pi\sim 2870$MHz for an NV. From Fig.~\ref{PCQ}(c), for $B_z\sim$ several gauss the selection rules $\Delta m_s=\pm 1$, hold and $\delta \nu_\pm/\delta B_z\sim \pm 28$ GHz/T.
\begin{figure}[h]
\begin{center}
\setlength{\unitlength}{1cm}
\begin{picture}(4,4)
\put(-1.3,-.1){\includegraphics[width=7cm,height=4cm]{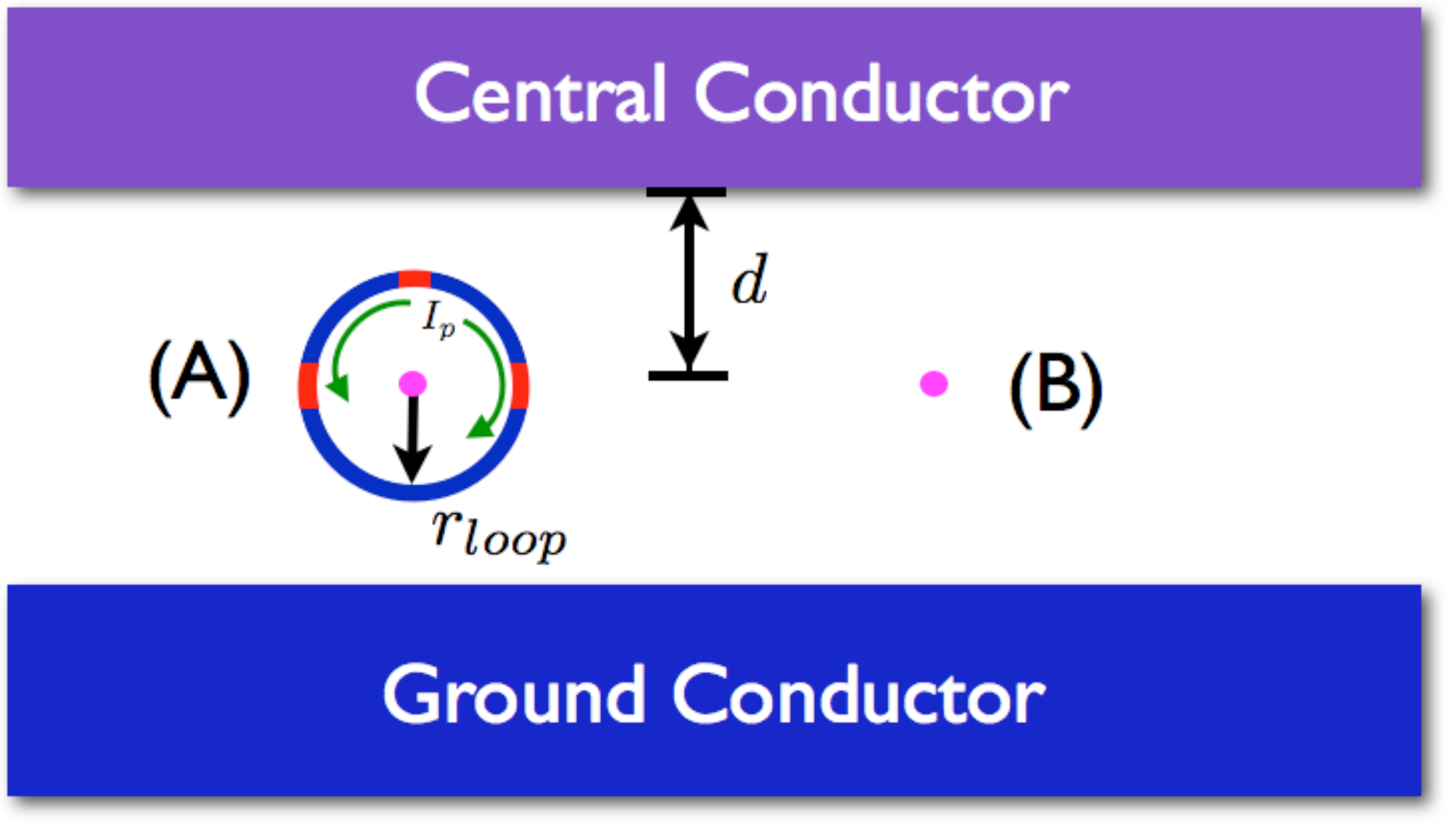}}
\end{picture}
\end{center}
	\caption{\label{Comparison}\textbf{Comparison of magnetic coupling strengths of the magnetic spin system (magenta), to the resonator}, (A) via a circular looped PCQ of radius $r_{loop}$, or (B) directly to the coplanar resonator.}
	\end{figure}
Let us consider now an NV placed a distance $d=50$nm away from the central conductor of the CPW resonator where $B_{0,rms}\sim 2.5$ milligauss. It will couple magnetically via the Zeeman term in (\ref{NV-Ham}), and using (\ref{B0-CPW}), the size of this coupling will be $|\bar{g}|/2\pi\sim 2.5\times 10^{-7}\times 28$ GHz$\sim 7$ kHz, while for $d\sim 5\mu$m, we have $|\bar{g}|/2\pi\sim 70$Hz. These couplings are far below the linewidth of the best CPW resonators fabricated to date and thus the direct magnetic coupling to a single NV so close to the resonator will not be resolved. 
Ideally, the ability to strongly couple to a single electronic spin system would allow for far easier quantum control of the coupled light-matter system but the tiny size of the CPW-single NV when directly coupled gives one little hope that this could be possible. In the following sections we describe how this {\em is possible}, by encircling the magnetic spin system with a Persistent Current Qubit (PCQ).

%\subsection{Persistent current qubit (PCQ)} 
{Persistent current qubit (PCQ):-}  A persistent current qubit (PCQ) \cite{Orlando:1999p7777}, is formed when a superconducting loop is interrupted by three Josephson junctions (Fig \ref{PCQ}), where all junctions are identical except that one is smaller by a factor $\alpha>0.5$, than the other two. When the loop is biased by a magnetic flux which is close to half a flux quantum, the device is an effective two level system \cite{Mooij:1999p7749}, with the qubit made up of two counter-circulating persistent currents. The effective two level system (or PCQ), is described by the Hamiltonian $\hat{H}_q=\frac{\hbar}{2}(\epsilon \hat{\sigma}_z+\Delta \hat{\sigma}_x)$, with $\epsilon=\frac{2I_p}{\hbar}(\Phi_x-\Phi_0/2)$, and $\Phi_x$ is the external magnetic flux through the loop. Going to an eigenbasis we can write $\hat{H}_q=\frac{\hbar}{2}\omega_0\hat{\sigma}_z$, with $\omega_0=\sqrt{\Delta^2+\epsilon^2}$. Recent work \cite{Niemczyk:2009p7767}, has $\alpha=0.7$, and $I_p\sim 450$nA, while $\Delta/2\pi\sim 5.2$ GHz. Persistent currents as large as $I_p\sim 800$nA have been observed \cite{Paauw:2009p7722}, while the area of PCQ loops are typically $A\sim 1-2\mu {\rm m}^2$.

{\em Coupling of the NV indirectly to the Coplanar Resonator via the PCQ:-} The strong coupling of a coplanar resonator to a PCQ has been recently demonstrated \cite{Abdumalikov:2008p7742}.
To estimate the coupling strength we note  $\hat{H}_{CPW-PCQ}=-\hat{\vec{\mu}}\cdot\hat{\vec{B}}$, where $\hat{\vec{\mu}}$ is the magnetic dipole of the PCQ induced by the circulating persistent currents of magnitude $I_p$,  $|\hat{\vec{\mu}}|=I_pA$, and where $A$ is the area of the PCQ loop. From (\ref{B0-CPW}), for a PCQ a distance $d$ from the central CPW conductor we find
\begin{equation}
|g|\sim \frac{I_pA\mu_0I_{rms}}{\hbar \pi d}=\frac{I_p\mu_0}{\hbar}\left(\frac{r_{loop}^2}{d}\right)\sqrt{\frac{\hbar\omega_r}{2L_r}}\;\;,
\label{gPCQ}
\end{equation}
where we have assumed a circular PCQ loop of radius $r_{loop}$. Taking $r_{loop}=0.8\mu$m, $I_p=600$nA and $L_r=2$ nH, we get $|g|/2\pi\sim 28.7$MHz. The Hamiltonian describing this coupling in the case where $\omega_0\sim\omega_r$, can be written as $\hat{H}_{CPW-PCQ}=\hbar g(\hat{a}^\dagger\hat{\sigma}^-+\hat{a}\hat{\sigma}^+)$, where $\hat{a}$ destroys a photon in the CPW while $\hat{\sigma}^-$ excites the qubit states of the PCQ.
%, $\hat{\sigma}^- |k\rangle_{PCQ}=(1-k)|k\oplus 1\rangle_{PCQ}$. 

We now consider placing the circular PCQ loop around an NV so that the NV is at the centre of the loop. As has been noted previously \cite{Orlando:1999p7777}, the persistent currents present in a PCQ generate sizable changes in magnetic flux within the loop $\Delta \Phi\sim 10^{-3}\Phi_0$.  Typically one surrounds the PCQ with a sensitive SQUID detector to measure the PCQ qubit via these small flux changes. In what follows we use the PCQ (without the SQUID), as a magnetic interconnect, coupling the NV through to the CPW resonator. We first note that the PCQ must be nominally biased by a magnetic flux $\Phi=\Phi_0/2$ to operate in the
regime where the states corresponding to  counter circulating currents
are degenerate. This yields a static $B_{s}\sim\Phi_0A/2$, magnetic field at the centre of the loop and $B_s\sim 5$ gauss for $A=2\;\mu{\rm m}^2$. We now estimate the small changes in magnetic field at the centre of the loop generated by the persistent counter-circulating currents, and from these, the small changes in the NV transition frequencies as these alter the Zeeman term in the NV's Hamiltonian. The magnetic field at the center of the loop due to  the persistent currents $I_p$, is $\vec{B}_{I_p}=\pm 2\mu_0 A I_p/(4\pi r_{loop}^3)\hat{z}$.
Further, exact placement of the NV at the centre of the loop is not required as the induced magnetic field varies slowly there.  
This magnetic field leads to a small shift in the NV's microwave transition frequencies ($m_s\rightarrow \pm 1$), of $\eta/4\pi \equiv\Delta \nu\sim \pm |\vec{B}_{I_p}|\times 28$ GHz/T. Obviously as one reduces $r_{loop}$ while retaining relatively large persistent currents $I_p$, one can increase $\eta$. Through this small change in magnetic field the PCQ qubit state can thus couple to the NV through the Zeeman term and we can now write the full NV Hamiltonian with the coupling to the PCQ as 
\begin{eqnarray}
\hat{H}_{NV-PCQ}&=&\frac{1}{2}\hbar\eta\hat{\sigma}_z\hat{S}_z+\hbar g_e\beta_eB_s\hat{S}_z\nonumber\\
&+&\hbar D\left(\hat{S}_z^2-\frac{2}{3}\mathbb{I}\right)\label{zfs}
\end{eqnarray}
 where $\hat{\sigma}_z$, the PCQ Pauli $z-$operator, couples directly to the NV triplet $\hat{S}_z$ operator.

%\subsection{Full Hamiltonian}
{\em Full Hamiltonian:-}
Using the above we are now able to describe the Hamiltonian of the coupled CPW-PCQ-NV system as
\begin{eqnarray}
\hat{H}&=&\hbar \omega_r\left(\hat{a}^\dagger \hat{a}+\frac{1}{2}\right)+\hbar\frac{\omega_0}{2}\hat{\sigma}_z+\hbar g_e\beta_e B_s\hat{S}_z+ {\rm ZFS}\nonumber\\
&+&\hbar g\left(\hat{a}^\dagger\hat{\sigma}^-+\hat{a}\hat{\sigma}^+\right)\nonumber\\
&+&\hbar\frac{\eta}{2}\hat{\sigma}_z\hat{S}_z\\
&+&\hbar\zeta \left(e^{-i\omega t}\hat{a}^\dagger+e^{i\omega t}\hat{a}\right)\nonumber\;\;,\label{bigHam}
\end{eqnarray}
where $ZFS$ is zero field splitting (second line), of (\ref{zfs}), and where we have included a term which drives the CPW resonator at rate $\zeta$. Driving the cavity resonantly, $\omega=\omega_r$, we can move to an interaction picture defined by the first line in (\ref{bigHam}), with $\omega_0 \sim \omega_r$, to find
\begin{eqnarray}
\hat{H}_I&=&\hbar\frac{\delta}{2}\hat{\sigma}_z+\hbar\zeta \left(\hat{a}+\hat{a}^\dagger\right)\nonumber\\
&+&\hbar g\left(\hat{a}^\dagger\hat{\sigma}^-+\hat{a}\hat{\sigma}^+\right)+\hbar\frac{\eta}{2}\hat{\sigma}_z\hat{S}_z+H_{decay}\;\;,\label{IntHam}
\end{eqnarray}
where the detuning between the PCQ and CPW resonator is $\delta=\omega_0-\omega_r$, and where $H_{decay}$, (which we model more specifically below), denotes decay and dephasing from the cavity, PCQ and NV. 

From the above analysis it is clear that as one reduces the size of the PCQ loop the coupling to the CPW decreases while the coupling to the NV increases. In Fig \ref{couplings}, we plot the dependence of the couplings as a function of loop radius and persistent current. From this we obtain the central result of this paper: that if one can fabricate PCQ loops with $r_{loop}\sim 0.4\mu$m (or smaller), and $I_p\sim 600$nA (or larger), then the couplings $[g_{CPW-PCQ},g_{NV-PCQ}, g_{NV-CPW}]/2\pi\sim[14{\rm MHz},60{\rm kHz},1{\rm kHz}]$, while $\kappa/2\pi\sim26$kHz \cite{Niemczyk:2009p7767}. This indicates that the NV-PCQ coupling will be resolvable through the spectroscopy of the CPW and that the {\em\bf NV will be effectively  strongly coupled through the PCQ interconnect into the stripline resonator.}

To examine how this coupling alters when we include realistic decay models, we write the full phenomenological quantum Master equation $\dot{\hat{\rho}}={\cal L}\hat{\rho}=-i[\hat{H}_I,\hat{\rho}]+\bar{{\cal L}}\hat{\rho}$, where 
\begin{equation}
\bar{{\cal L}}\hat{\rho}=\sum_{j=1}^5\,\left[\hat{C}_j\hat{\rho}\hat{C}_j^\dagger-\frac{1}{2}\left\{\hat{C}_j^\dagger\hat{C}_j,\hat{\rho}\right\}\right]\;\;,
\label{master}
\end{equation}
and $\hat{C}_j=\{\sqrt{\kappa}\hat{a}, \sqrt{\gamma_{PCQ}}\hat{\sigma}^+, \sqrt{\gamma_{NV}}\hat{S}^+, \sqrt{\gamma_{\phi\,PCQ}}\hat{\sigma}_z,$ $\sqrt{\gamma_{\phi\, NV}}\hat{S}_z]$, where we have damping of the  CPW resonator at rate $\kappa$, decay of the PCQ/NV qubits $\gamma_{PCQ/NV}/2\pi=1/T_{1\, PCQ/NV}$, and their associated dephasing rates $\gamma_{\phi\, PCQ/NV}/2\pi=1/T_{\phi\,PCQ/NV}=1/T_{2\,PCQ/NV}-1/2T_{1\,PCQ/NV}$. We compute the power spectrum of the cavity under the small driving $\zeta$, from
\begin{equation}
S(\omega)=\frac{1}{2\pi}\int_{-\infty}^\infty\;e^{-i\omega\tau}\langle \hat{a}^\dagger(\tau+t)\hat{a}(t)\rangle\,d\tau\;\;,
\label{power}
\end{equation}
where we use the quantum regression theorem $\langle \hat{a}^\dagger(\tau+t)\hat{a}(t)\rangle={\rm Tr}[\hat{a}^\dagger e^{{\cal L}\tau}\hat{a}\hat{\rho}_{ss}]$, where $\hat{\rho}_{ss}$, is the steady state of the Master equation (\ref{master}). With just the CPW coupled to the PCQ we expect to see a very large vacuume Rabi splitting and these peaks will be further slightly split due to the interaction of the PCQ with the NV.
Flux qubits fashioned to-date suffer from relatively large decay and dephasing rates $T_{1\,PCQ}=10T_{2\,PCQ}=2\mu{\rm s}$. However these rates might be lowered by engineering the devices in a more symmetric layout as proposed by \cite{Steffen:2009p7738}, and with this in mind we take the following decoherence parameters for our simulations: $\{\kappa/2\pi,T_{1\,PCQ},T_{1\,NV},T_{2\,PCQ},T_{2\,NV}\}=[26{\rm kHz}, 20\mu{\rm s}, 4000\mu{\rm s}, 2\mu{\rm s}, 600\mu{\rm s}]$. In Fig \ref{splittings}, we plot $S(\omega)(r_{loop})$, and note that the NV splitting in the resonator spectrum is quite sensitive to the dephasing.

{\em Multi-Turn interconnects:-} In the previous section we showed that one can amplify the magnetic coupling of the NV through to the CPW resonator by encompassing the NV with a PCQ circular loop. The coupling between the NV and PQC increased with decreasing loop radius but this also decreases the coupling between the PCQ and the resonator. There may also be technical difficulties in fabricating very small PCQ structures. To circumvent this one can consider multi-turn PCQ loops (Fig \ref{multiturn}), where the circular loop of the PCQ winds multiple times around the NV, thus scaling up the resulting $B$-field induced by the PCQ circulating currents $I_p$, and thus scaling up the strength of the PCQ-NV coupling, and also the PCQ-CPW coupling. Such a structure may require a free air-bridge. Using multi-turn interconnects may allow one much more flexibility to strengthen the effective coupling strength between the NV electronic spin system and the CPW resonator (see Fig. \ref{dephasing} and Fig \ref{dephasing_equal}). However this may come at the expense of shorter dephasing times for the PCQ system as its couplings to any stray magnetic fluctuators will also be amplified.

{\em Summary:-} Circuit-QED has already demonstrated strong coupling between solid-state qubits and a superconducting bus and this heralds a route towards the future construction of large scale quantum devices. Through our PCQ interconnect one has the potential to strongly couple individual, long lived, electronic or perhaps, nuclear, spins into the superconducting bus. This will allow one to use such systems in quantum devices for information processing or metrology as long lived quantum memories,  or to deterministically entangle individual atomic solid-state systems over centimeter length scales, or, in the case of an individual NV, optically readout and reset the quantum state of the system.

\begin{figure*}[!ht]
\begin{center}
\setlength{\unitlength}{1cm}
\begin{picture}(18,4.5)
\put(0,0){\includegraphics[width=6.2cm,height=4.8cm]{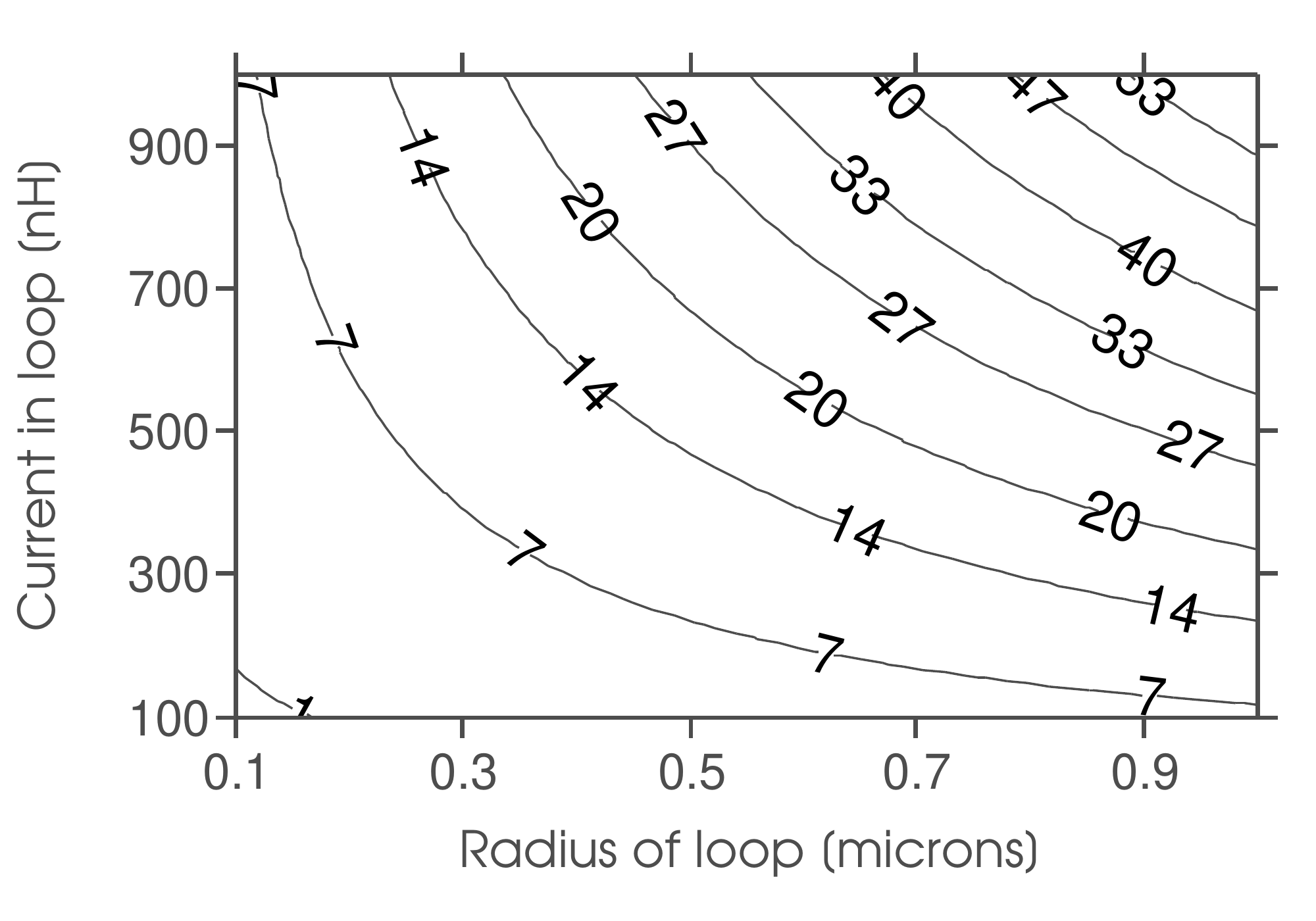}}
\put(6,0){\includegraphics[width=6.2cm,height=4.8cm]{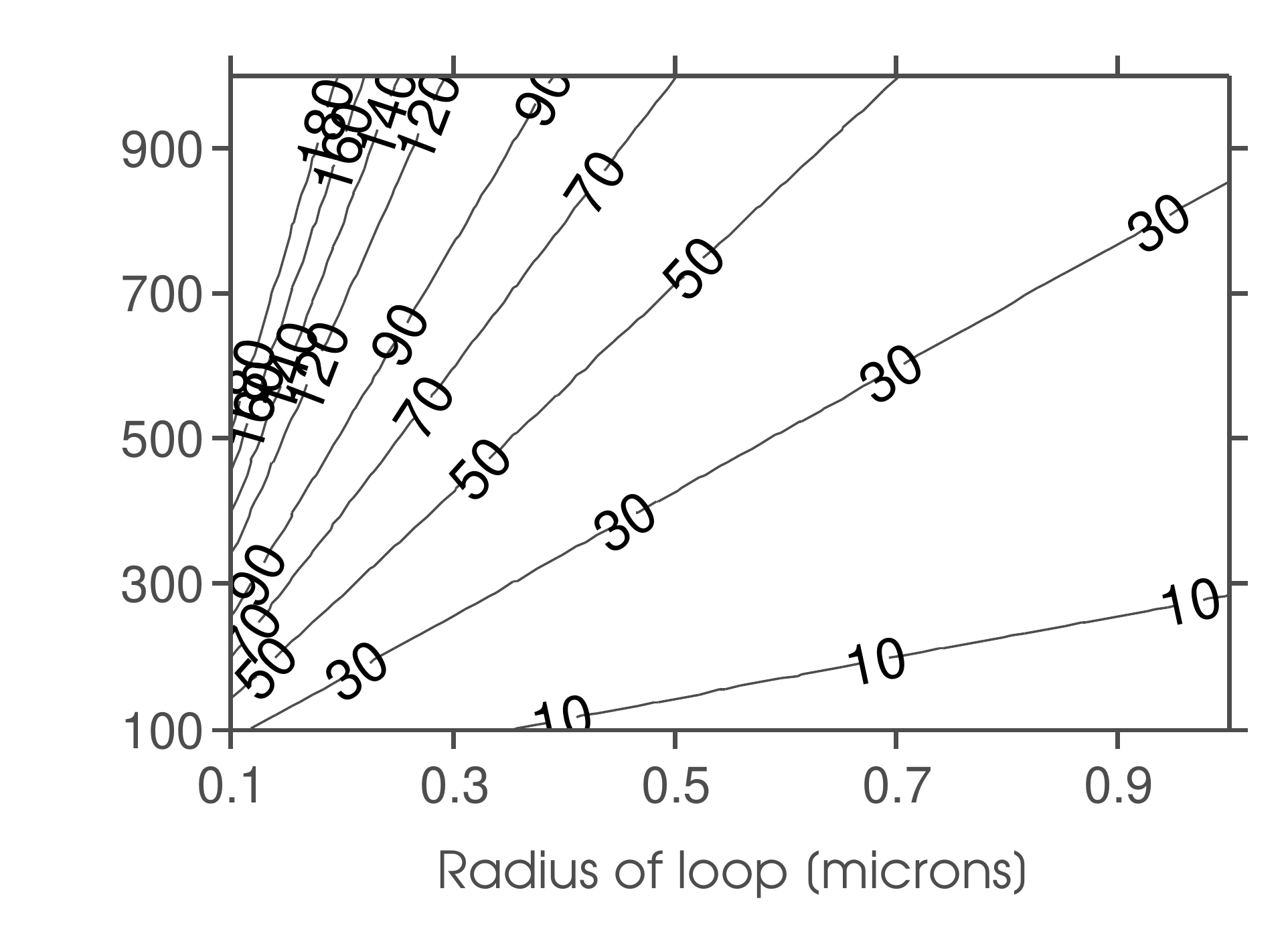}}
\put(12,0){\includegraphics[width=6.2cm,height=4.8cm]{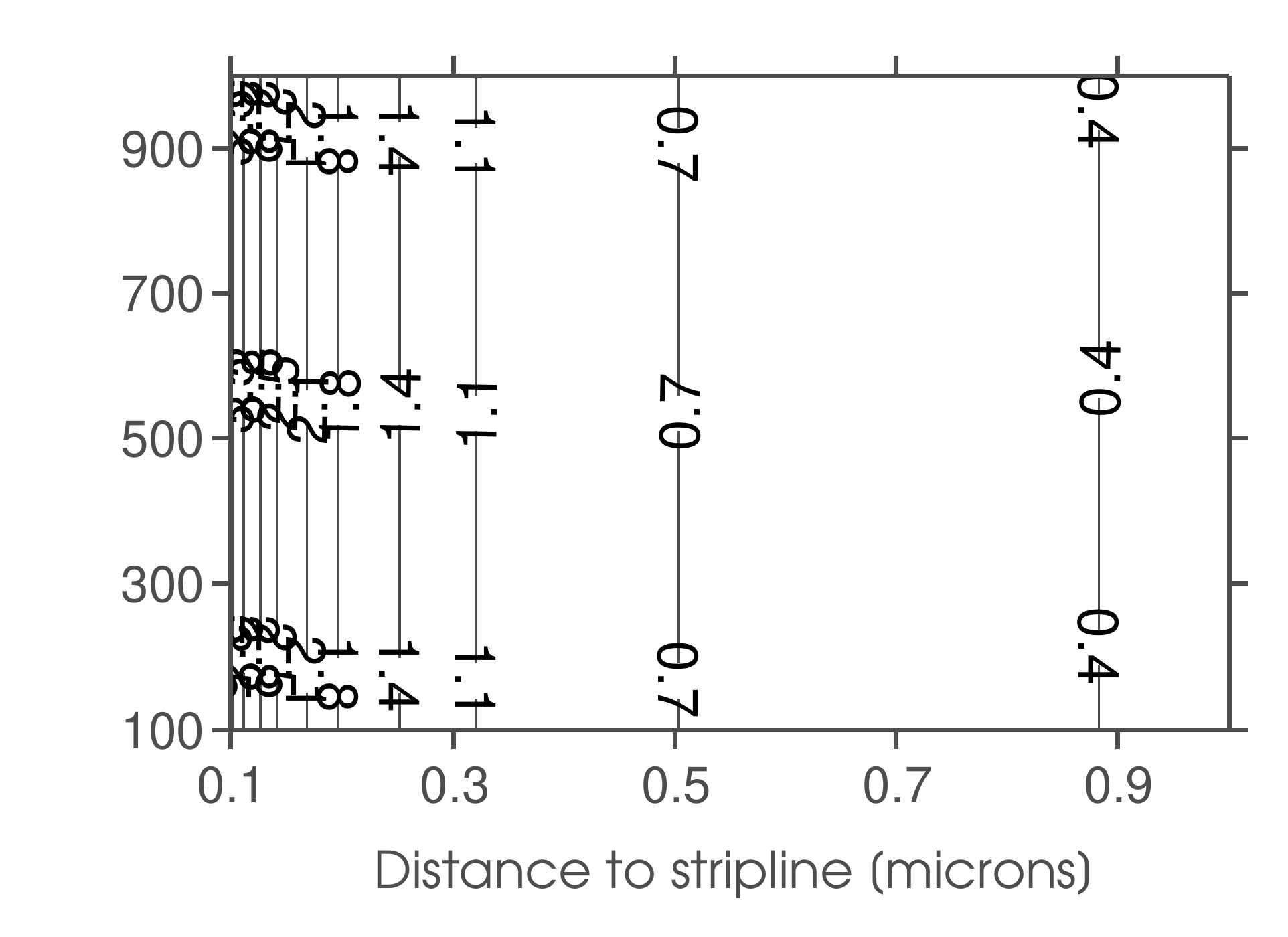}}
\put(0,0){\large(A)}
\put(6.1,0){\large(B)}
\put(12,0){\large(C)}
\end{picture}
\end{center}
\caption{\textbf{Comparison of coupling strengths:} (A) strength of coupling between the persistent current qubit and the coplanar resonator as a function of $r_{loop}$ and $I_p$, i.e. $g/2\pi$, in MHz; (B) coupling between the NV and PCQ, i.e.  $\eta/2\pi$, in kHz; (C) direct coupling between the NV and CPW in kHz.}
\label{couplings}
\end{figure*}

\begin{figure}[h]
\begin{center}
\setlength{\unitlength}{1cm}
\begin{picture}(4,11)
\put(-1.9,4.2){\includegraphics[width=8cm,height=5.5cm]{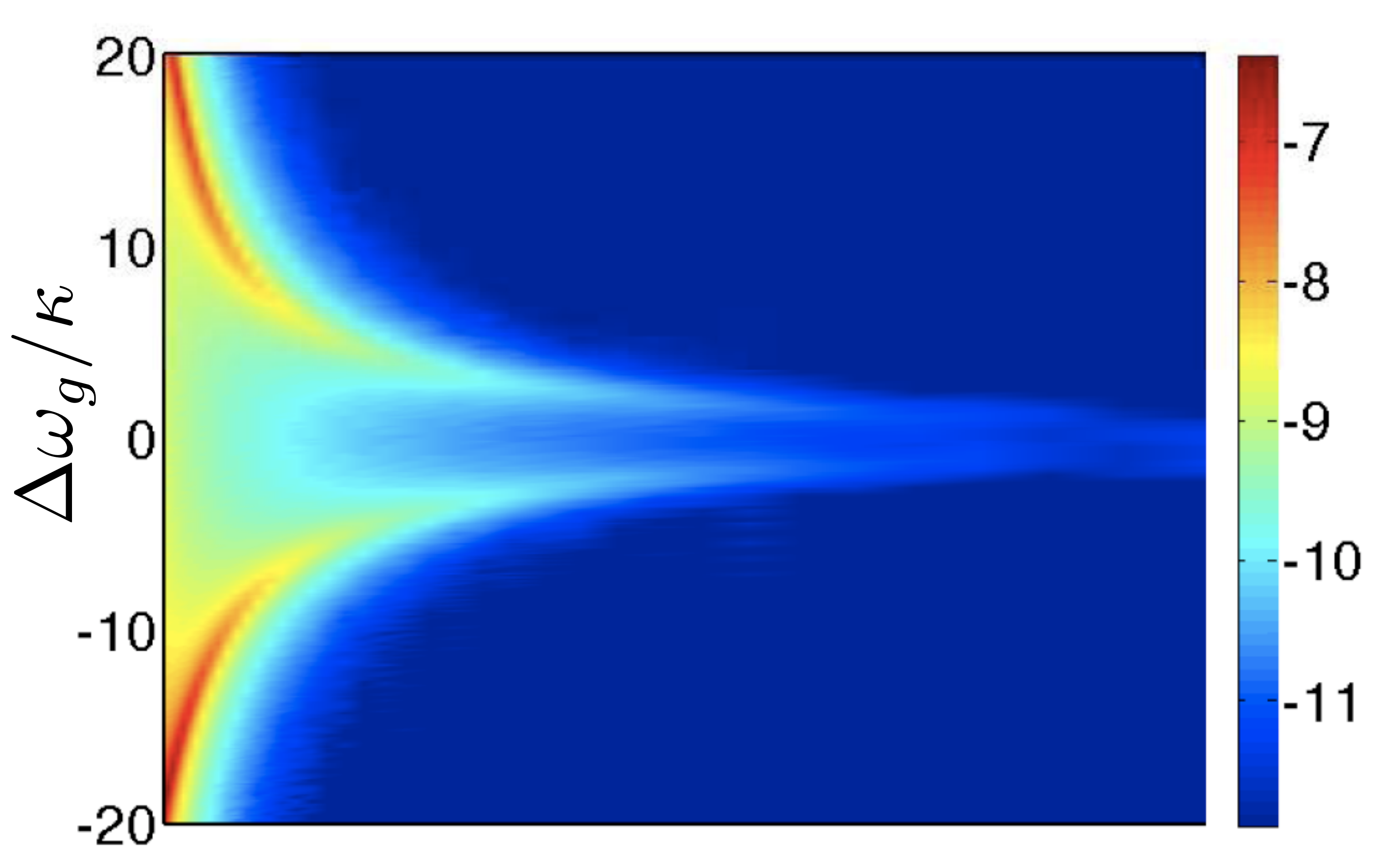}}
\put(-2.,-.3){\includegraphics[width=8cm,height=5.cm]{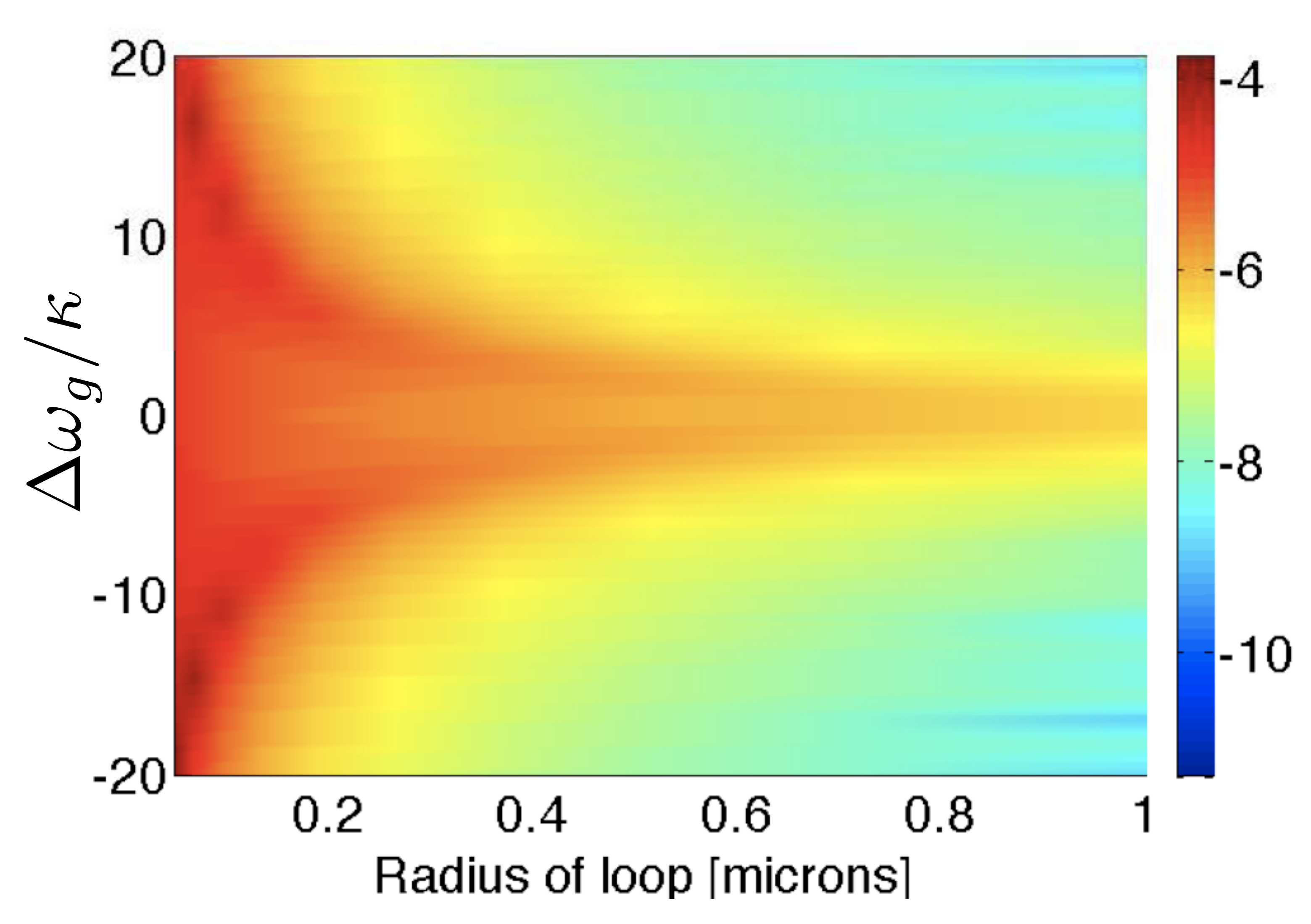}}
\put(-2.1,9.3){\large(A)}
\put(-2.1,4.1){\large(B)}
\end{picture}
\end{center}
\caption{\textbf{Splitting of the cavity spectrum due to the NV}. Steady state power spectrum of the cavity $S(\omega)$, as a function of the PCQ loop radius. We consider only one of the vacuum Rabi peaks for $\delta=0$, centered at $\omega=\omega_g\equiv g$, and plot $\log_{10}[S(\Delta \omega_g)]$, where $\Delta\omega_g\equiv \omega-\omega_g$. We choose $\zeta=2\kappa,\;\;I_p=800$nA, $T_{1\,NV}=4$ms, $T_{2\,NV}=600\mu$s, $\kappa/2\pi=26$ kHz, $\omega_0=\omega_r=2\pi\times 6$ GHz and $E_p=2\kappa$.  (A) We omit the dephasing terms in (\ref{master}), and this corresponds to the case when $T_{2\,PCQ}=2T_{1\,PCQ}$;  (B) we set the dephasing to be moderately large, $T_{2\,PCQ}=T_{1\,PCQ}$.}
\label{splittings}
\end{figure}

\begin{figure}[h]
\begin{center}
\setlength{\unitlength}{1cm}
\begin{picture}(7,7)
\put(-1,0){\includegraphics[width=9cm,height=7cm]{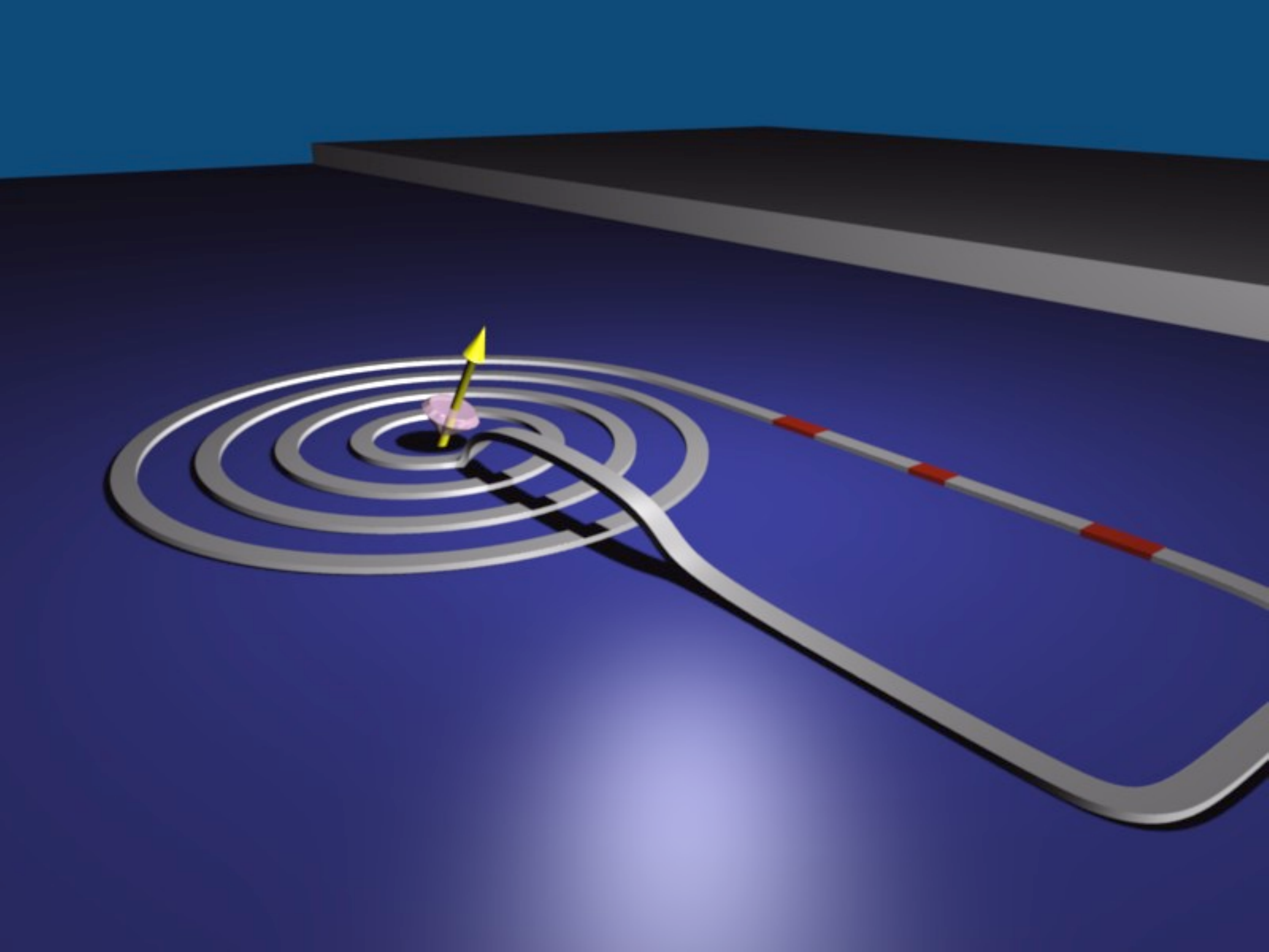}}
\end{picture}
\end{center}
\caption{\textbf{Mutli-turn PCQ interconnect} By creating an $n-$looped spiral inductor incorporating the three Josephson junction PCQ one can amplify the PCQ coupling strengths to the resonator (distant and in grey), and the NV (diamond pink) $n$-times.}
\label{multiturn}
\end{figure}

\begin{figure}[h]
\begin{center}
\setlength{\unitlength}{1cm}
\begin{picture}(4,11)
\put(-2.,4.2){\includegraphics[width=8cm,height=5.cm]{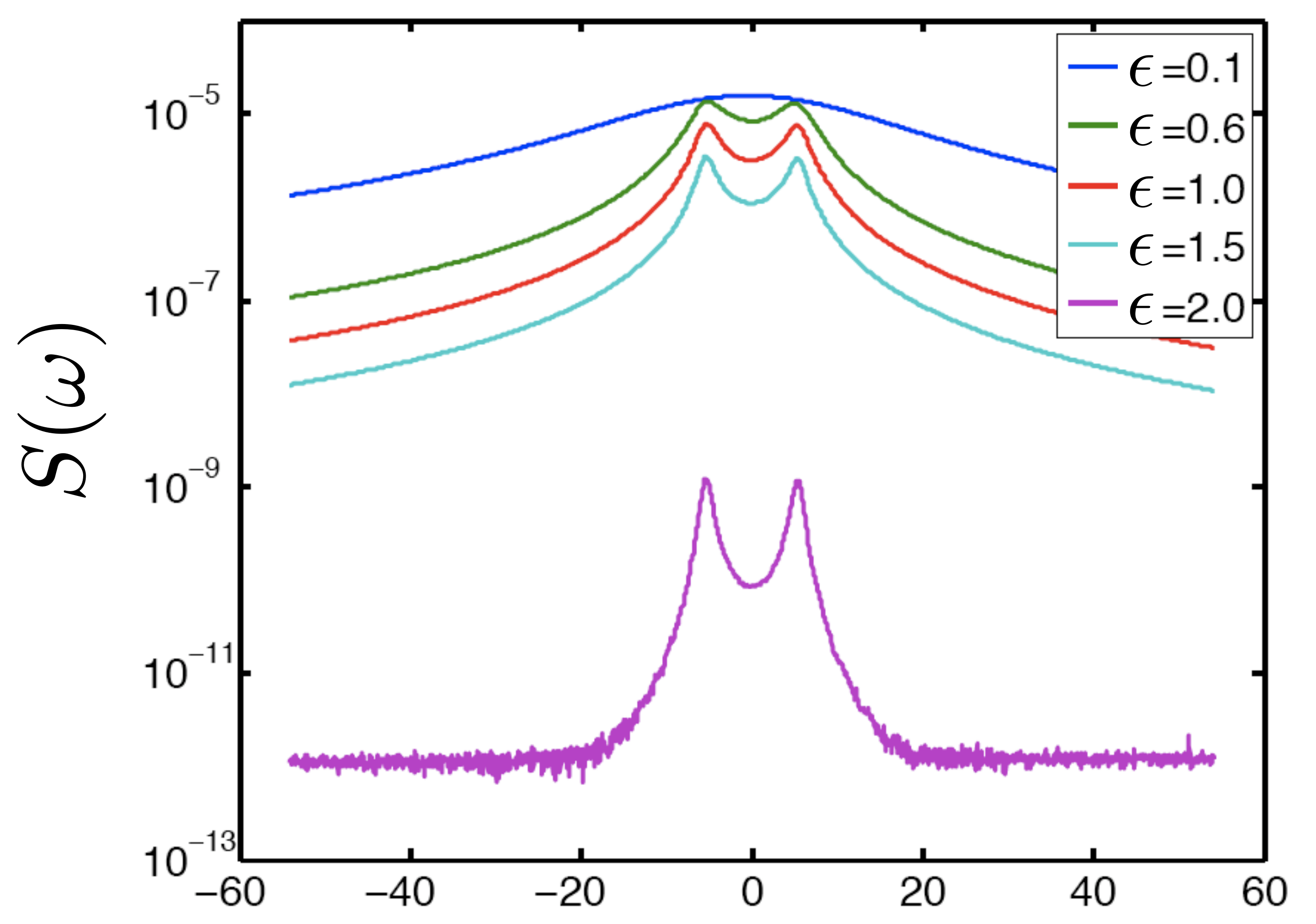}}
\put(-2.,-.5){\includegraphics[width=8cm,height=5.cm]{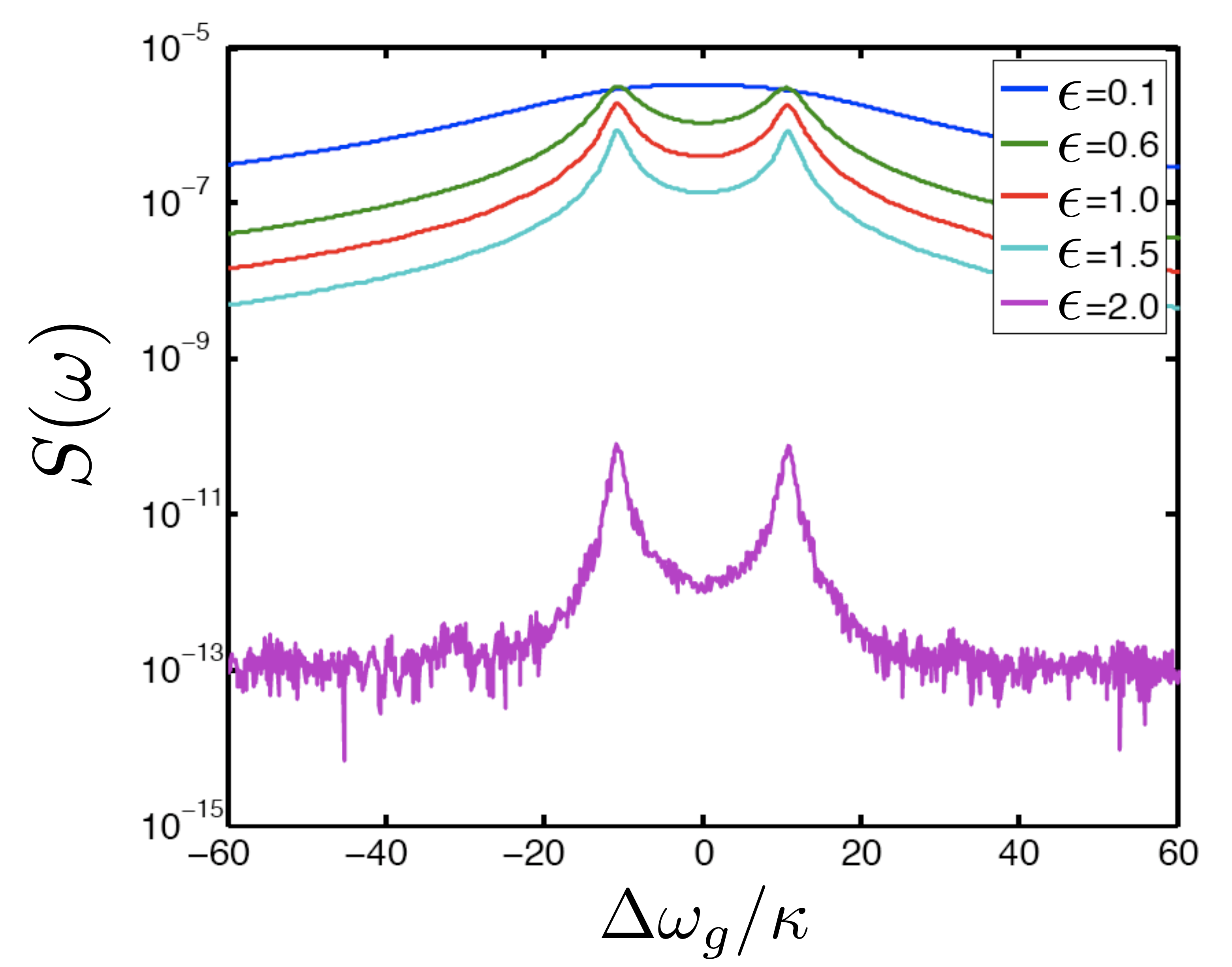}}
\put(-2.1,9.3){\large(A)}
\put(-2.1,4.1){\large(B)}
\end{picture}
\end{center}
\caption{\textbf{Dependence of the NV splitting of the PCQ Vacuume Rabi line  on $T_{2\,PCQ}$} for a single turn and a multiturn PCQ qubit. In (A) we choose $T_{1\,PCQ}=20\mu$s, and set $T_{2\,PCQ}=\epsilon T_{1\,PCQ}$. We take $I_p=880$nA and $r_{loop}=0.2	\mu$m and the other parameters as in Fig (\ref{splittings}). In (B) we consider a two turn loop PCQ. This doubles the CPW-PCQ and PCQ-NV coupling strengths but may reduce $T_{2\,PCQ}$. The lower curves in both graphs display small amounts of numerical roundoff noise.}
\label{dephasing}
\end{figure}

\begin{figure}[h]
\begin{center}
\setlength{\unitlength}{1cm}
\begin{picture}(4,8)
\put(-2.2,-.8){\includegraphics[width=9.5cm,height=7.cm]{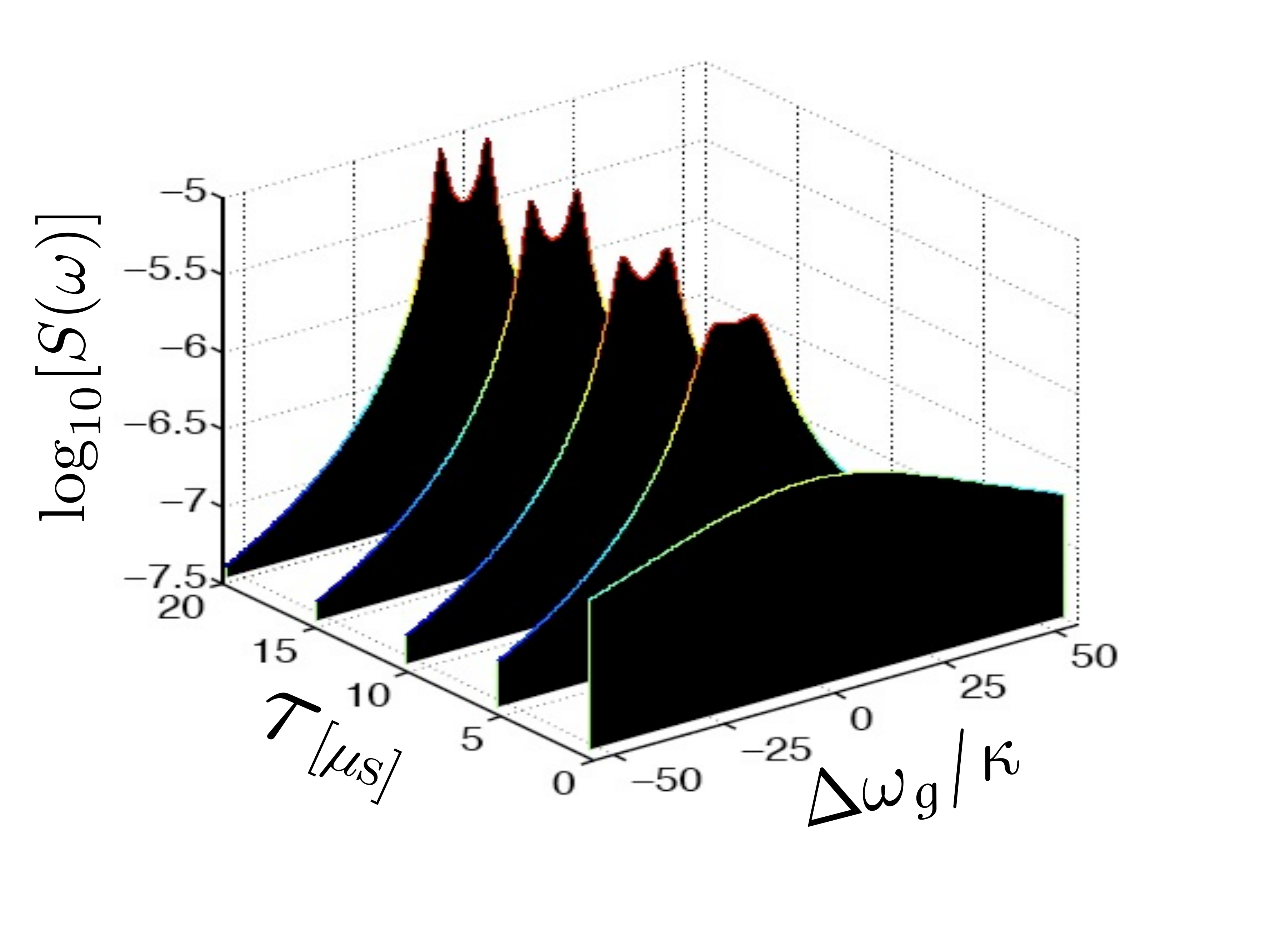}}
\end{picture}
\end{center}
\caption{Dependence of the NV splitting of the PCQ Vacuume Rabi line  on the decoherence rates. We plot $\log_{10}[S(\omega)]$, with the parameters as in Fig. (\ref{dephasing}), but set $T_{1\,PCQ}=T_{2\,PCQ}=\tau=[0.5,5.0,10.0,15.0,20.0]\mu$s. We see that we would require $\tau>5\mu$s to begin to resolve the splitting.}
\label{dephasing_equal}
\end{figure}

%\footnotesize
%\begin{multicols}{2}
\bibliographystyle{Science}
\bibliography{Nov1Twamley_1}
%\end{multicols}

\end{document}